\documentclass[3p,fleqn]{elsarticle}
\usepackage{balance}
\usepackage{graphicx}
\usepackage{longtable}
\usepackage{wrapfig}
\usepackage{amssymb,amsmath,amsthm}
\usepackage[colorlinks=true,
    linkcolor=blue,
    filecolor=magenta,
    urlcolor=cyan]{hyperref}

\usepackage[mathscr]{euscript}
\usepackage{upgreek}
\usepackage[symbol]{footmisc}
\usepackage{slashed}
\usepackage{cancel}
\usepackage{tocloft}
\usepackage{float}
\usepackage{setspace}
\usepackage[normalem]{ulem}
\usepackage[figuresright]{rotating}
\biboptions{sort&compress}

\newcommand{\mpl}{M_\text{Pl}}

\usepackage{geometry}
\numberwithin{equation}{section}

\usepackage{fancyhdr}
\begin{document}

\begin{frontmatter}

\title{Extended Effective Field Theory of Dark Energy:\\Ghost Condensate Dark Energy with Sextic Dispersion Relation in de Sitter Spacetime}

\author[af1]{A. Ashoorioon\footnote{amjad@ipm.ir}}
 \author[af1,af2]{M. B. Jahani Poshteh\footnote{jahani@ipm.ir}}
\author[af1]{A. Yousefi-Sostani\footnote{yousefiae@ipm.ir}}
\address[af1]{School of Physics, The Institute for Research in Fundamental Sciences (IPM), P.O. Box 19395-5531, Tehran, Iran}
\address[af2]{Universit\`a di Camerino, Divisione di Fisica, Via Madonna delle carceri 9, 62032 Camerino, Italy}

\begin{abstract}

We continue our studies of the ghost condensate (GC) with sixth-order dispersion relation. Contrary to the GC with quartic dispersion relation, we find that the correction to the Newtonian potential explicitly depends on the space and time dependence of matter density. At late times when the Newtonian potential becomes time-independent, one obtains similar oscillatory behavior at the distance $\frac{\mpl}{M^2}$, but this time at the time scale $\frac{M^4}{\mpl^3}$, where $M^2$ is the ghost field velocity.
We also show that the speed of gravitational wave is modified in a frequency dependent manner at momenta close to $\frac{\mpl}{\sqrt{|\sigma_1|}}$, where $\sigma_1$ is the coefficient of $\gamma^{ij} \nabla_i K_{lr} \nabla_j K^{lr}$ operator in the unitary gauge action.

\end{abstract}
 \end{frontmatter}

\thispagestyle{fancy}

\numberwithin{equation}{section}


\thispagestyle{fancy}

\tableofcontents

\section{Introduction}
\label{intro}



Dark Energy (DE), an enigmatic component of the universe, is responsible for the recent observed acceleration of cosmic expansion. Its discovery in the late 1990s through observations of distant Type Ia supernovae \cite{SupernovaSearchTeam:1998fmf, SupernovaCosmologyProject:1998vns} fundamentally transformed our understanding of cosmology. It is now understood that the universe consists of roughly 70$\%$ DE, 25$\%$ dark matter, and only about 5$\%$ ordinary baryonic matter \cite{Planck:2018nkj}. Despite its dominance, the true nature of DE remains one of the most profound and unresolved mysteries in modern physics. In fact, the puzzle of DE underscores the interplay between physical cosmology and theoretical innovation. One of the most recent major challenges related to DE is the tension in measurements of the Hubble constant ($H_0$). Different methods for determining the universe's expansion rate yield conflicting results. Local measurements, such as those based on Cepheid-calibrated supernovae distances, give higher values of Hubble constant ($H_0 \approx 73 \ \textrm{km} \ \textrm{s}^{-1} \textrm{Mpc}^{-1}$ \cite{Riess:2021jrx}) compared to those inferred from the cosmic microwave background ($H_0 \approx 67 \ \textrm{km} \ \textrm{s}^{-1} \textrm{Mpc}^{-1}$ \cite{Planck:2018nkj}), i.e. a $4\sigma - 6\sigma$ mismatch. This discrepancy, often referred to as ``Hubble tension", may point to new physics, potentially linked to DE \cite{hubbleTension}.

The nature of DE is deeply entangled with several theoretical and observational challenges. One major issue is the fine-tuning problem. If DE is associated with the cosmological constant ($\Lambda$) in Einstein's field equations, its observed value is inexplicably small. Quantum field theory predicts a vacuum energy density that is $10^{122}$ times larger than what is observed (the so-called cosmological constant problem \cite{Weinberg:1988cp, Carroll:1991mt, Lombriser:2019jia}). This enormous discrepancy raises fundamental questions about why the cosmological constant is so finely tuned to its current value. Another issue is the coincidence problem \cite{Zlatev:1998tr,Steinhardt:1999nw,Wang:1999fa}. Observations indicate that the energy density of DE and dark matter are of the same order of magnitude today, despite their vastly different evolutionary dynamics over time ($\rho_\Lambda \propto {\rm const.}$ vs. $\rho_\textrm{DM} \propto a^{-3}$, where $a$ is the scale factor). This suggests a puzzling synchrony, as cosmic acceleration began only recently ($z \sim 0.3$, where $z$ is redshift). This raises the question of why the universe's acceleration began relatively recently in cosmic history, a coincidence that remains unexplained.

To address the challenges presented by DE, numerous theoretical models have been proposed, ranging from the straightforward cosmological constant to more complex frameworks involving dynamic fields or modifications of gravity. Each idea offers unique perspectives, though none fully resolves the mysteries of DE.

As mentioned before, one of the simplest explanations for DE is the cosmological constant $\Lambda$. It represents a constant energy density permeating space uniformly, aligning well with observational data. Despite its simplicity, the cosmological constant fails to satisfactorily address the fine-tuning and coincidence problems. It also provides no insight into why the universe's accelerated expansion began only in relatively recent cosmic history, leaving critical questions unanswered. Maybe the simplest next step is dynamical cosmological constant. This approach aims to address the cosmological constant problem by allowing the vacuum energy density to change, potentially offering a more natural explanation for the observed acceleration of the universe's expansion \cite{10.1093/mnras/251.1.128, Mukohyama:2003nw, Yoshimura:2022iog, Sola:2016zeg}. Such a DE is further characterized by its equation of state parameter. Observations suggest that $w \approx -1$ \cite{Planck:2018nkj, Riess:2021jrx}, almost consistent with a cosmological constant, but uncertainties leave room for alternative dynamical models where $w \neq -1$ \cite{10.1046/j.1365-8711.1999.02551.x, Sola:2006uk, Tripathi:2016slv, Janka:2022krt}. It remains a central goal of cosmological research to accurately constrain $w$. One framework to realize such a scenario is known as quintessence model, which introduces a dynamical scalar field denoted by $\phi$, which evolves over time. The field's energy density and pressure are determined by a potential $V(\phi)$, allowing for variations in the equation of state parameter $w$. This feature makes quintessence models more flexible than the cosmological constant, as they offer potential solutions to the coincidence problem. However, the dynamics of $\phi$ requires fine-tuning to match observations, and the fundamental origin of the scalar field remains mysterious \cite{Ratra:1987rm, Caldwell:1997ii, Tsujikawa:2013fta}. Further extensions of quintessence, such as k-essence and phantom model, incorporate non-standard kinetic terms in the scalar field Lagrangian. These models exhibit a broader range of behaviors, including phantom energy scenarios where $w < -1$. Phantom models predict a big rip scenario. While conceptually intriguing, these models often encounter theoretical instabilities and inconsistencies. \cite{Caldwell:1999ew, Armendariz-Picon:2000ulo, Gibbons:2003yj, Ludwick:2017tox}.

The possibility that DE is a manifestation of a breakdown in general relativity on cosmological scales has also been proposed. This approach considers the possibility that DE is not an independent energy form but instead signals the need for modifications to general relativity on large scales. Modifying general relativity to account for DE, however, presents significant challenges. Any proposed modification must remain consistent with stringent observational constraints, such as those from solar system tests, making this an extremely difficult theoretical problem to address. An examples of this approach are $f(R)$ theories, which replace the Ricci scalar $R$ in the Einstein-Hilbert action with a general function $f(R)$. These models can account for late-time acceleration but require careful tuning to avoid contradictions with local gravitational tests \cite{Nojiri:2003ft, Nojiri:2010wj, Nojiri:2017ncd, Capozziello:2005ku, Amendola:2006kh, Tsujikawa:2007xu}.

Similarly, brane-world models, such as the Dvali-Gabadadze-Porrati (DGP) model, posit that our universe exists as a four-dimensional brane within a higher-dimensional bulk. Gravity's leakage into extra dimensions at large distances can mimic DE effects, though such models face challenges in maintaining consistency with observational constraints \cite{Dvali:2000hr, Roos:2007yd, Lombriser:2009xg}. Also, massive gravity \cite{deRham:2010kj, deRham:2014zqa, Hinterbichler:2011tt} and its extended theories \cite{Hassan:2011zd} are other intriguing avenues, wherein the graviton is given a small mass, altering gravitational interactions on cosmological scales. While this approach is promising, it is burdened with significant theoretical and observational challenges, such as ensuring stability and compatibility with established physics \cite{Deser:2013gpa, Deser:2014fta}. However, efforts have been made to demonstrate that bimetric gravity theories can overcome observational and theoretical challenges by some considerations \cite{Hogas:2019ywm, Hogas:2021lns, Hogas:2021saw, Dwivedi:2024okk}.

Coupled DE models explore the idea of interaction between DE and dark matter or other cosmic components \cite{Pourtsidou:2013nha, Boehmer:2008av, Fay:2016yow}. These interactions could naturally address the coincidence problem by linking the evolution of DE and dark matter densities. However, these models introduce additional parameters and require precise calibration to remain consistent with observational data.

Emergent DE theories \cite{Li:2019yem, LozanoTorres:2024tnt, John:2023fsy, Yang:2020ope} suggest that DE arises as a byproduct of deeper physical phenomena, such as quantum effects, vacuum fluctuations, or entropy in a holographic framework. For example, the entropic gravity hypothesis posits that DE might result from the thermodynamic properties of spacetime itself \cite{Easson:2010av, Verlinde:2016toy, Diaz-Saldana:2018gxx}. These ideas shift the focus from DE as a fundamental entity to a secondary phenomenon rooted in other principles of physics.

Finally, ghost condensate models \cite{Arkani-Hamed:2003pdi, Mukohyama:2006be} offer a different perspective, proposing that a scalar field with an unconventional kinetic term condense into a stable vacuum configuration. This generates a modified low-energy effective theory with a Lorentz-violating dispersion relation (in the case of \cite{Arkani-Hamed:2003pdi} a quartic dispersion relation), where DE emerges dynamically. Ghost condensation can mimic the cosmological constant while allowing for richer behavior, including Lorentz-violating effects, and even phantom-like acceleration if coupled to dilaton \cite{Piazza:2004df}. Additionally, it provides testable observational predictions, such as deviations in the growth of cosmic structures or spin-dependent forces.

In principle, one could propose a GC theory with a sixth-order dispersion relation too. Nevertheless, the authors of \cite{Arkani-Hamed:2003pdi} argued that only GC theories that do not generate dispersion relations beyond the quartic order are acceptable as effective field theories. Their reason is that the interacting terms in theories with dispersion relations beyond quartic dispersion, relation become strong in the infrared limit. However, we showed in \cite{Ashoorioon:2023zju} that a sixth-order ghost condensate theory could still be used to describe the current cosmological acceleration of the universe, as the outgoing modes during the latest acceleration epoch have not gone too far outside the horizon. Below we will review the arguments briefly.

Let us consider the scalar field $\phi$ with the action
\begin{equation} \label{action-full} 
A= A_0+A_1\,,
\end{equation}
where
\begin{equation}
A_0 = M^4 \int \mathrm{d}^4 x \sqrt{-g} P(X)\,,\label{action0} 
\end{equation}
and
\begin{eqnarray} \label{action1} 
A_1=\int \mathrm{d}^4 x \sqrt{-g} \Big[ M^2(\tilde{q}(X)(\Box\phi)^2+\bar{q}(X)(\partial_{\mu}\partial_{\nu}\phi)^2) + \tilde{s}(X) \left( \partial_\mu \Box \phi \right)^2+ \bar{s}(X) \left( \partial_\mu \partial_\nu \partial_\rho \phi \right)^2 \Big]\,,
\end{eqnarray}
in which $X = \partial_{\mu}\phi\partial^{\mu}\phi$. \footnote{The spacetime signature is mostly negative $(+, -, -, -)$.}  In an arbitrary background, the equation of motion (EOM) derived from the action \eqref{action0}
takes the form,
\begin{equation}\label{eom-Px} 
\partial_\mu [\partial^\mu \phi P'(X) ]=0\,.
\end{equation}
A Lorentz-violating solution is given by
\begin{equation} \label{LVsolution} 
\phi = ct\,,
\end{equation}
where $c$ is a constant. If a small fluctuation is considered around this solution \eqref{LVsolution}, such that
\begin{equation} \label{pi-introducing} 
\phi = c t + \pi\,,
\end{equation}
then $P(X)$ can be expanded around it, resulting in
\begin{align} \label{action0-taylor-expansion} 
A_0 = & M^4 \int \mathrm{d}^4 x \sqrt{-g} [ \dot{\pi}^2 ( P'(c^2) + 2 c^2 P''(c^2) ) - P'(c^2) (\partial_i \pi)^2 - 2 P''(c^2) c \dot{\pi} (\partial_i \pi)^2+\cdot\cdot\cdot]\,,
\end{align}
subject to the following conditions
\begin{equation} \label{stability-condition} 
P'(c^2) > 0\,, \text{\hspace{0.5cm}} P'(c^2) + 2 c^2 P''(c^2) > 0\,.
\end{equation}
The latest conditions ensure that kinetic and gradient terms appear with proper signs. In an expanding background, $P'(\dot{\phi}^2)\propto a^{-3}$ and thus the coefficient of $(\partial_i\pi)^2$ vanishes for $\dot{\phi}^2=c_{\ast}^2$. Therefore, when one considers the term proportional to $\tilde{q}$ and $\bar{q}$, it leads to $\omega^2\sim k^4$ dispersion relation in the absence of gravity. Therefore, in the absence of gravity, the stabilizing terms such as $(\partial_{\mu}\phi\partial^{\mu}\phi)^2$ lead to a non-Lorentzian dispersion relation
\begin{equation} \label{disperion-relation1} 
\omega^2 \sim \frac{\bar{M}^2}{M^4} k^4+\frac{\bar{S}}{M^4}k^6\,,
\end{equation}
where
\begin{align} \label{M2bar} 
\bar{M}^2 &=-2 M^2 (\tilde{q}(c_*^2)+\bar{q}(c_*^2))\,,\\
\bar{S} & =-2(\tilde{s}(c_*^2)+\bar{s}(c_*^2))\,. \label{Sbar} 
\end{align}
To obtain a pure sixth order dispersion relation we have to either assume $\bar{M}^2=0$ or $\frac{\bar{M}^2}{\bar{S}}\ll k^2\sim \frac{H^2}{(1+z_c)^2}$.

The argument based on which the sixth and higher order dispersion relations were ruled out as a viable effective field theory, is based on how the interacting operators like
\begin{equation} \label{interacting operator} 
\int \mathrm{d}^3 x \mathrm{d}t M^4 \dot{\pi} \left(\partial_i \pi \right)^2,
\end{equation}
behaves under the energy scaling $E\rightarrow \mathsf{S} E$. Assuming that the dispersion relation is sixth order, such an interacting operator behaves like $\mathsf{S}^{-1/3}$ and therefore the theory becomes strongly coupled in
the IR. However, in the late-time acceleration, which is like a low-scale inflation, the largest mode that has exited the horizon is $\sim \frac{H}{1+z_c}$, where $z_c$ is the redshift of domination of dark energy. Like inflation, the
onset of strong coupling could be defined as when
  \begin{equation}\label{strong-coupling} 
  |f_{\rm {}_{NL}}|\ll \zeta^{-1}\,,
  \end{equation}
  where $\zeta=-H\pi$, and
  \begin{equation} \label{fnl-def} 
  |f_{\rm {}_{NL}}|\sim \frac{B(k_1,k_2,k_3)}{\Delta_{\zeta}^4(k)}\, ,
  \end{equation}
in which $\Delta_{\zeta}^2(k)=\frac{k^3}{2\pi^2}\langle\zeta(k) \zeta(k)\rangle$ and $B(k_1,k_2,k_3)=\langle\zeta(k_1) \zeta(k_2)\zeta(k_3)\rangle$. The inequality \eqref{strong-coupling}, for the largest modes $k_1=k_2=k_3=H/(1+z_c)$ and in the limit of $H/M\ll 1$ reduces to
\begin{equation}\label{cSbar-relation} 
   c_* \bar{S}^{1/4}\gg 10^{-34}\,,
\end{equation}
which is satisfied for $c_*$ and $\bar{S}$ of $\mathcal{O}(1)$.  $|f_{\rm {}_{NL}}|$ becomes comparable to  $\zeta^{-1}$, only about $77$ e-folds from now, which is far in the future.

In the next section, we investigate the action for the ghost condensate in the unitary gauge in an expanding background and find out how the gravitational potential is modified at large distances. We also obtain the equation for the dark energy perturbations. In the third section, we obtain how the speed of gravitational waves is modified in a GC with sixth order dispersion relation. Finally we conclude our paper.

\section{Sextic Ghost Condensate Coupled to Gravity}

In this section, it is beneficial to modify our language for simplicity. Instead of coupling the Lagrangian for $\phi$ to gravity directly, it is suitable to go to the ``unitary gauge" that makes the  modifications of gravity manifest. The unitary gauge eliminates $\phi$ as a degree of freedom, but preserves residual gauge freedom associated with time-dependent spatial diffeomorphisms. By expanding the action around flat space, we can construct quadratic terms that remain invariant under diffeomorphisms. However, to introduce a genuine modification of general relativity (GR), higher-order terms are necessary to render the field Golsdtone boson of time-diffeomorphism, $\pi$, dynamical. For this purpose, we can construct invariants involving higher derivatives, such as $K_{ij}$ (and $\nabla_i K_{jk}$), which denotes the linearized extrinsic curvature of constant time surfaces (and its derivative) within the theory. The $K_{ij}^2$ terms have been invoked and used for the ghost condensate with quadratic dispersion relation before \cite{Arkani-Hamed:2003pdi}. As it was argued in \cite{Ashoorioon:2023zju}, and reviewed in the previous section, it is still plausible to consider ghost condensates with higher order, and in particular sixth order, dispersion relation.

Noting that $K_{ij} \rightarrow  K_{ij}+ \partial_{ij} \pi $ under the Stueckelberg trick, to realize the sixth order dispersion relation, we modify the unitary gauge action as follows,
  \begin{align}
      A_\text{GC} = \int \mathrm{d}^4x \frac{\sqrt{\gamma}}{2} \Big[ \frac{M^4}{4}(X-1)^2 - \gamma^{ij} \left( \sigma_1 \nabla_i K_{lr} \nabla_j K^{lr} +\sigma_2 \nabla_i K\nabla_j K \right)
      -\sigma_3\nabla_{i}K^{i}_{j}\nabla_{l}K^{lj} -\sigma_4 \nabla^{i}K_{ij}\nabla^{j}K \Big]\,,\label{AGC} 
  \end{align}
where $\sigma_{i}$'s are constants following \cite{Ashoorioon:2018uey}, and $K = K^i_i$. We first employ the perturbed metric in the Newtonian gauge
\begin{equation}\label{newtonian-metric} 
\mathrm{d}s^2 = (1+ 2\Phi(x^\mu))\mathrm{d}t^2 - (1-2 \Psi(x^\mu))\delta_{i j} \mathrm{d}x^i \mathrm{d}x^j\,,
\end{equation}
where $\Phi$ and $\Psi$ are two scalar potentials which characterize the gravitational perturbations. Assuming the absence of the anisotropic inertia terms and in the Newtonian limit, one finds that $\Phi = \Psi$. Including
the ghost condensate $\pi$, working in the Newtonian limit $\omega^2\ll k^2$, and going to the canonical normalization for $\Phi$ and $\pi$, one obtains the dispersion relation \cite{Ashoorioon:2023zju}
\begin{equation}\label{k6GC-dispersion-relation} 
\omega ^2 = \frac{S_1}{M^4}\left(k^6 - \mu^2 k^4 \right)\,,
\end{equation}
where $S_1 \equiv \sum_{i=1}^{4}\sigma_i$ and $\mu^2 \equiv M^4/2 M^2_\text{Pl}$. It is noticed that the $k^4$ term is induced due to the mixing with gravity while its sign is opposite to the $k^6$ term. It illustrates that $\omega^2$ gets negative values for $k < \mu$, i.e. an instability  like the Jeans
 instability has appeared. The largest negative value of $\omega^2$ is
\begin{equation}\label{upsilon-def} 
  \omega_\text{ins}^2 = - \frac{2 S_1}{27} \frac{\mu^4}{M_\text{Pl}^2} \equiv - \Upsilon^2,
\end{equation}
which introduces a timescale $t_c \sim \Upsilon^{-1}$ for the instability, or when the deviation from GR should happen. It is worth noting that $\mu^{-1}$ characterizes the length scale $r_c$ over which the modifications to the Newtonian potential occurs.

To investigate the present accelerating expansion of the universe, one has to depart from the Minkowski  to the de Sitter background. The perturbed metric becomes
\begin{equation}\label{ds-metric} 
  \mathrm{d}s^2 = (1+ 2\Phi(x^\mu))\mathrm{d}t^2 - a^2(t) (1-2 \Psi(x^\mu))\delta_{i j} \mathrm{d}x^i \mathrm{d}x^j\,,
\end{equation}
where $a(t) = e^{H t}$ while $H = {\rm constant}$. In addition to the Einstein-Hilbert action
\begin{equation}\label{EH-action} 
    A_\text{EH} = \frac{1}{2}\int \mathrm{d}^4 x \sqrt{-g} M_\text{Pl}^2 R\,,
\end{equation}
we have the action \eqref{AGC}, whose expression in the terms of $\Phi$, $\Psi$, and $\pi$ takes the form
\begin{align}
 \mathcal{L}_\text{GC} = - \frac{1}{2 a^3} \Big[ & \sigma_1 \big[  \partial_i  (\partial_j^2 \pi) + H a^2 \partial_i \left(\Phi - \dot{\pi}\right) + a^2 \partial_i \dot{\Psi}  \big]^2 +
 \sigma_2 \big[ \partial_i  (\partial_j^2 \pi) + 3 H a^2 \partial_i (\Phi - \dot{\pi}) +3 a^2 \partial_i \dot{\Psi} \big]^2 \nonumber \\ & + \sigma_3 \big[ \partial_i  (\partial_j^2 \pi) +
 a^2 \partial_i \dot{\Psi} \big]^2 + \sigma_4 \big[ \big( \partial_i  (\partial_j^2 \pi) + 2 H a^2 \partial_i (\Phi - \dot{\pi}) + 2 a^2 \partial_i \dot{\Psi} \big)^2 \nonumber \\ &
 - \big( H a^2 \partial_i (\Phi - \dot{\pi}) + a^2 \partial_i \dot{\Psi} \big)^2\big]\Big] + \frac{1}{2} M^4 a^3 \left( \Phi - \dot{\pi} \right)^2\ . \label{LGC} 
\end{align}
Here, the variation of action \eqref{AGC} plus the Einstein-Hilbert action with respect to the inverse of the metric yields the EOM as follows
\begin{align}
 \mpl^2 G^{\rho}_{\lambda} - \frac{1}{2} M^4 \delta^\rho_0 \delta_{\lambda 0 } (1-X) -\frac{1}{2}\left(\sigma_1 \mathcal{K}_{{}_1 \lambda}^{\ \rho} + \sigma_2 \mathcal{K}_{{}_2 \lambda}^{\ \rho} + \sigma_3 \mathcal{K}_{{}_3 \lambda}^{\ \rho} + \sigma_4 \mathcal{K}_{{}_4 \lambda}^{\ \rho}\right)=0\, , \label{EOM-EH-GC} 
\end{align}
where $G^\rho_\lambda$ stands for the Einstein tensor and
\begin{align}
   \mathcal{K}_{{}_1 \lambda}^{\ \rho} = & \left(-\frac{1}{2} \delta^\rho_\lambda \gamma^{\mu \nu}  + \gamma^{\mu \rho} \gamma^\nu_\lambda\right) \nabla_\mu K^{\alpha \beta} \nabla_\nu K_{\alpha \beta} + \frac{1}{2} \nabla_\sigma  \nabla_\mu \left(\gamma^{\mu \nu} \nabla_\nu K_{\alpha \beta}\right)\left[ \gamma^{\sigma \alpha} \gamma^{\beta}_{\lambda} n^\rho + \gamma^{\sigma \beta} \gamma^{\alpha}_{\lambda} n^\rho - \gamma^{\alpha \rho} \gamma^\beta_\lambda n^\sigma \right] \nonumber \\ &  + \frac{1}{2} \nabla_\mu \left(\gamma^{\mu \nu} \nabla_\nu K_{\alpha \beta}\right)\left[ K^{\alpha \beta}n^\rho n_\lambda -n_\lambda \left( n^\alpha K^{\beta \rho} + n^\beta K^{\alpha \rho} \right) + K \left( \gamma^{\alpha \rho} \gamma^{\beta}_{\lambda} - \gamma^{\alpha \rho} \delta^{\beta}_{\lambda} - \gamma^{\beta}_{\lambda} g^{\alpha \rho} \right) \right. \nonumber \\ & \left. + \gamma^\beta_\lambda \nabla^\alpha n^\rho + \gamma^{\alpha \rho} \nabla^\beta n_\lambda  - 4 K^\alpha_\lambda g^{\beta \rho} \right] + \nabla_\mu \left(\gamma^{\mu \nu} \nabla_\nu K^\rho_{\beta} \right) K^\beta_\lambda + \gamma^{\mu \nu} \nabla_\mu K^\beta_\lambda \nabla_\nu K^\rho_\beta + \nabla_\alpha \left( \gamma^{\nu \rho} \nabla_\nu K_{\lambda \beta} \right) K^{\alpha \beta}  \nonumber \\ & + \gamma^{\nu\rho} \nabla_\nu K_{\lambda \beta} \nabla_\alpha K^{\alpha \beta} - \nabla_\alpha \left( \gamma^{\nu \rho} \nabla_\nu K_{\alpha \beta} \right) K^\beta_\lambda - \gamma^{\nu\rho} \nabla_\nu K^{\alpha}_{\beta} \nabla_\alpha K^\beta_\lambda + \frac{1}{2}\nabla_\sigma \nabla_\nu\left( \gamma^{\mu \nu} \nabla_\mu K^{\alpha \beta} \right) \left[ \gamma^{\sigma}_{\alpha} \gamma_{\beta\lambda} n^\rho \right. \nonumber \\ & \left. + \gamma^{\rho}_{\alpha} \gamma^\sigma_\nu n_\lambda - \gamma^{\rho}_{\alpha} \gamma_{\beta \lambda}n^\sigma \right]  + \frac{1}{2} \nabla_\nu\left( \gamma^{\mu \nu} \nabla_\mu K^{\alpha \beta} \right) \left[ K_{\alpha \beta}n^\rho n_\lambda -n_\lambda \left( n_\alpha K_{\beta}^{\rho} + n_\beta K_{\alpha}^{ \rho} \right) \right. \nonumber \\ & \left. + K \left( \gamma_{\alpha}^{ \rho} \gamma_{\beta \lambda} - \gamma_{\alpha}^{\rho} g_{\beta\lambda} - \gamma_{\beta\lambda} \delta_{\alpha}^{\rho} \right) + \gamma_{\beta\lambda} \nabla_\alpha n^\rho + \gamma_{\alpha}^{\rho} \nabla_\beta n_\lambda \right] - \nabla_\alpha\left( \gamma^\mu_\lambda \nabla_\mu K^{\alpha\beta}\right)K^\rho_\beta - \gamma^\mu_\lambda \nabla_\mu K^{\alpha \beta} \nabla_\alpha K^\rho_\alpha
    \nonumber \\ & - \nabla_\mu \left( \gamma^{\mu\nu} \nabla_\nu K^{\alpha\rho}\right)K_{\lambda\alpha}  - \gamma^{\mu\nu} \nabla_\nu K^{\alpha \rho} \nabla_\mu K_{\lambda\alpha} + \nabla_\alpha \left(\gamma^\mu_\lambda\nabla_\mu K^{\beta \rho}\right) K^{\alpha}_\beta + \gamma^\mu_\lambda\nabla_\mu \nabla_\alpha  K^{\beta \rho} K^{\alpha}_\beta\,, \label{k1}
\\
    \mathcal{K}_{{}_2 \lambda}^{\ \rho} = & \left(-\frac{1}{2} \delta^\rho_\lambda \gamma^{\mu\nu} + \gamma^{\mu\rho} \gamma^\nu_\lambda \right) \nabla_\mu K \nabla_\nu K + \nabla_\nu (\gamma^{\mu\nu}\nabla_\mu K) \left[ 2 K^\rho_\lambda + K\left(n^\rho n_\lambda + \gamma^\rho_\lambda \right) - \nabla^\rho n_\lambda - \nabla_\lambda n^\rho \right] \nonumber \\ & -\nabla_\alpha \nabla_\nu (\gamma^{\mu\nu}\nabla_\mu K)\left[ \gamma^{\alpha \rho} n_\lambda + \gamma^\alpha_\lambda n^\rho - \gamma^\rho_\lambda n^\alpha \right]\,, \label{k2}
\\
    \mathcal{K}_{{}_3 \lambda}^{\ \rho} = & -\frac{1}{2} \delta^\rho_\lambda \nabla_\mu K^\mu_\nu \nabla_\alpha K^{\alpha \nu} - \frac{1}{2} \nabla_\beta \nabla_\mu \nabla_\alpha K^{\alpha \nu}\left[ \gamma^{\beta \mu} \gamma^{\beta}_{\lambda} n^\rho + \gamma^{\sigma \beta} \gamma^{\alpha}_{\lambda} n^\rho - \gamma^{\alpha \rho} \gamma^\beta_\lambda n^\sigma \right]   \nonumber \\ & + \frac{1}{2} \nabla_\mu \nabla_\alpha K^{\alpha \nu}\left[ K^{\mu}_\nu n^\rho n_\lambda -n_\lambda \left( n^\mu K^{\rho}_\nu + n_\nu K^{\mu \rho} \right) + K \left( \gamma^{\mu \rho} \gamma_{\nu \lambda} - \gamma^{\mu \rho} g_{\nu\lambda} - \gamma^{\mu}_{\lambda} \delta^{\rho}_\nu \right) + \gamma_{\nu\lambda} \nabla^\mu n^\rho + \gamma^{\mu \rho} \nabla_\nu n_\lambda \right. \nonumber \\ & \left. -  K^\mu_\lambda \delta_{\nu}^{\rho} \right] -\frac{1}{2} \nabla_\alpha \nabla_\beta \nabla_\mu K^\mu_\nu \left[ \gamma^{\alpha \beta} \gamma^\nu_\lambda n^\rho + \gamma^{\alpha \rho} \gamma^{\beta \nu} n_\lambda - \gamma^{\beta \rho} \gamma^{\nu}_{\lambda} n^\alpha \right] + \frac{1}{2} \nabla_\mu \nabla_\alpha K^\mu_\nu \left[ K^{\alpha \nu} n^\rho n_\lambda \right. \nonumber \\ & \left. -n_\lambda \left( n^\alpha K^{\nu \rho} + n^\nu K^{\alpha \rho} \right) + K \left( \gamma^{\alpha \rho} \gamma^{\nu}_{\lambda} - \gamma^{\alpha \rho} \delta^{\nu}_{\lambda} - \gamma^{\mu}_{\lambda} g^{\alpha \rho} \right) + \gamma^\nu_\lambda \nabla^\alpha n^\rho + \gamma^{\alpha \rho} \nabla^\mu n_\lambda  - 4 K^\alpha_\lambda g^{\nu \rho} \right] \nonumber \\ & + \frac{1}{2} \delta^\rho_\lambda \nabla_\alpha \left(K^{\alpha \nu} \nabla_\mu K^{\mu}_{\nu} + K^\alpha_\nu \nabla_\mu K^{\mu\nu} \right) - \nabla_\mu \left(K^\rho_\lambda \nabla_\nu K^{\mu \nu}\right) + \nabla_\mu \left( K^\mu_\lambda \nabla_\nu K^{\nu \rho}  \right)\,, \label{k3}
\\
    \mathcal{K}_{{}_4 \lambda}^{\ \rho} = & -\frac{1}{2} \delta^\rho_\lambda \nabla^\mu K_{\mu \nu} \nabla^\nu K + \nabla^\rho K_{\lambda \mu} \nabla^\mu K +\frac{1}{2}\nabla_\alpha \nabla^\mu \nabla^\nu K \left[\gamma^{\alpha}_\mu \gamma_{\nu\lambda} n^\rho + \gamma_{\mu}^\rho \gamma^\alpha_\nu n_\lambda - \gamma^\rho_\mu \gamma^\alpha \gamma_\nu n_\lambda - \gamma_\mu^\rho \gamma_{\nu\lambda} n^\alpha  \right] \nonumber \\ & +\frac{1}{2} \nabla^\mu \nabla^\nu K \left[ K_{\mu \nu}n^\rho n_\lambda -n_\lambda \left( n_\mu K_{\nu}^{\rho} + n_\nu K_{\mu}^{ \rho} \right) + K \left( \gamma_{\mu}^{ \rho} \gamma_{\nu \lambda} - \gamma_{\mu}^{\rho} g_{\nu\lambda} - \gamma_{\nu\lambda} \delta_{\mu}^{\rho} \right) + \gamma_{\nu\lambda} \nabla_\mu n^\rho + \gamma_{\mu}^{\rho} \nabla_\nu n_\lambda \right] \nonumber \\ & - \nabla_\lambda \left( K^{\mu\rho} \nabla_\mu K \right) -\frac{1}{2} \left(\nabla^\mu \nabla^\nu K_{\mu\nu}\right) K \left( 2 \gamma^\rho_\lambda - \delta^\rho_\lambda \right) + \frac{1}{2} \nabla_\alpha \nabla^\mu \nabla^\nu K_{\mu\nu} \left( \gamma^\alpha_\lambda n^\rho + \gamma^{\alpha \rho} n_\lambda - \gamma^\rho_\lambda n^\alpha \right)\,.
\end{align}
With several integrations by parts and defining new variables
\begin{align}
   S_2 \equiv 3 \sigma_1 + \sigma_2 + 2 \sigma_4\,,
    \quad S_3 \equiv \ 3 (\sigma_1 + 3 \sigma_2 + \sigma_4)\,, \ \ \quad S_4 \equiv  2 S_2 + \sigma_3\,, \label{Si-def} 
\end{align}
and varying w.r.t. $\pi$ one obtains the following equation
\begin{align}
  \left( \frac{\partial}{a} \right)^6 \pi -\frac{S_2}{S_1}H^2\left( \frac{\partial}{a} \right)^4 \pi + \frac{S_4}{S_1} H \left( \frac{\partial}{a} \right)^4 \Phi +
   \frac{2\mu^2 M_\text{Pl}^2 }{S_1} \Big[3H (\dot{\pi} - \Phi) + ( \ddot{\pi} - \dot{\Phi}) \Big] = 0 \,. \label{EOM-pi} 
\end{align}
One can divide equation \eqref{EOM-EH-GC} into three equations by the components: $00$, $0i$, and $ij$. By utilizing these three equations along with Eq. \eqref{EOM-pi}, we can derive separate equations for the gravitational potentials and $\pi$, independent of one another. The off-diagonal components of \eqref{EOM-EH-GC} suggest that $\Psi = \Phi$ \footnote{In fact the off-diagonal components of \eqref{EOM-EH-GC} yields
\begin{equation}\label{phi-psi-general}
   \left(\frac{\nabla}{a}\right)^2 (\Phi-\Psi)=-\frac{9}{2}H^2 (\Phi-\Psi)\,,
\end{equation}
which even in absence of anisotropic stress have solutions other than $\Psi = \Phi$. In this work we, however, focus on $\Psi = \Phi$.
}.

Finally, the equation which is obtained for $\pi$ is
\begin{align}
\left(\frac{\partial}{a}\right)^6 & \pi \ -\  C_1 \left(\frac{\partial}{a}\right)^6 \dot{\pi} \ + \  C_2\left(\frac{\partial}{a}\right)^4 \pi \ + C_3 \left(\frac{\partial}{a}\right)^4 \dot{\pi} + C_4 \left(\frac{\partial}{a}\right)^2 {\pi} \ + \ C_5\left(\frac{\partial}{a}\right)^2 \dot{\pi} \ + \ C_6 \left(\frac{\partial}{a}\right)^2  \ddot{\pi} \nonumber \\ & + C_7 \ddot{\pi} \ - \ C_8 \dot{\pi} =0\,, \label{pure-equation-pi}  
\end{align}
where
\begin{eqnarray}
C_1 & = & \frac{3 S_2 S_4 H}{12 S_1 \mpl^2 -3 S_2 S_4 H^2 }\, , \nonumber \\
C_2 & = &  \frac{S_2 \left(S_4 \left(\mu ^4-3 H^2 \mu ^2\right)-4 \mpl^2 \left(3 H^2+\mu ^2\right)\right)-3 S_3 S_4 H^4 }{12 S_1 \mpl^2 -3 S_2 S_4 H^2 }\, , \nonumber \\
C_3 & = & \frac{S_2 \left(3 H^2 S_4 \left(\mu ^2-3 H^2\right)+2 \mu ^2 \mpl^2\right)-3  S_3 S_4H^4 }{3 \left( S_2 S_4 H^3- 4 S_1 H \mpl^2 \right)}\, ,\nonumber\\
C_4 & = & \frac{3 S_3 S_4 H^4 \mu ^2 +4 S_3 H^2 \mu ^2 \mpl^2  - H^2 \mu ^4 S_3 S_4}{3 S_2 S_4 H^2 -12 S_1 \mpl^2 }\, , \nonumber \\
C_5 & = & \frac{S_3 \left(S_4 \left(9 H^5-3 H^3 \mu ^2\right)-2H \mu ^2 \mpl^2 \right)+12  S_4H \mu ^2 \mpl^2}{12 S_1 \mpl^2 -3  S_2 S_4 H^2}\, , \nonumber \\
C_6 & = & \frac{36  S_4 \mu ^2 \mpl^2}{4  S_1 \mpl^2- S_2 S_4 H^2 }\, , \nonumber 
\end{eqnarray}
\begin{eqnarray}
C_7 & = & \frac{4 \mu ^2 \mpl^2 \left(9 S_4 H^2 \left(\mu ^2-3 H^2\right)-2 \mpl^2 \left(H^2-3 \mu ^2\right)\right)}{S_2 S_4 H^4 -4  S_1 H^2 \mpl^2}\, , \nonumber \\
C_8 & = & 4 \mu ^2 \mpl^2  \Bigg(\frac{6 \mpl^2 \left(H^2-3 \mu ^2\right)}{S_2 S_4 H^3 -4 S_1 H \mpl^2} + \frac{ S_4 \left(30 H^4-19 H^2 \mu ^2+3 \mu ^4\right)}{ S_2 S_4 H^3 -4 S_1 H \mpl^2} \Bigg) \, .   \label{Ci-def}
\end{eqnarray}
Due to the order of magnitude discrepancy between $\mu$, $H$ and $\mpl$, the expressions for $C_i$'s imply that $k^6$ dominates in a cosmological background. Also, the sound speed will be defined as
\begin{equation}\label{cs} 
c_s^2 = \frac{C_7}{C_4} = \frac{ 12 \mpl^2 \left(9S_4 H^2 \left( \mu^2 - 3 H^2 \right) + 2 \mpl^2 \left( 3 \mu^2 - H^2 \right) \right)}{S_3 H^4 \left( S_4 \left(3 H^2-\mu ^2\right)+4 \mpl^2 \right)}\,.
\end{equation}

Now, we simply need to find the equation for $\Phi$ as its solution provides deviations from standard gravity in the low energy limit. The equation for $\Phi$ is obtained to be
\begin{align}
& \frac{S_1 S_3}{6 S_2 \mpl^2 \mu^2}  \left(\frac{\partial}{a}\right)^6 \Phi -\frac{S_1 S_3}{2S_2}\frac{H}{\mpl^2 \mu^2}\left(\frac{\partial}{a}\right)^4 \dot{\Phi} + \left[ \frac{2S_1}{3S_2 H^2} + \frac{ S_3 H^2}{6 \mpl^2 \mu^2}
\left(\frac{3 S_1}{S_2} - 1\right) -\frac{S_4}{6 \mpl^2}  \right]\left(\frac{\partial}{a}\right)^4 \Phi \nonumber \\ & + \left[\frac{S_3 H^4}{2\mpl^2\mu^2} - \frac{2S_1}{3S_2} \frac{\mu^2}{H^2} - \left( 1 + \frac{S_1}{S_2} \right)
\right] \left(\frac{\partial}{a}\right)^2 \Phi  +\left[\frac{S_3 H^3}{2\mpl^2\mu^2} - \frac{1}{3H} \right] \left(\frac{\partial}{a}\right)^2 \dot{\Phi} \nonumber \\ &+ 3 H^2 \Phi +  4 H \dot{\Phi} + \ddot{\Phi} = 0\,. \label{equation-phi} 
\end{align}
Besides the $k^6$ term, $k^4$ and $k^2$ has also appeared in the EOM for $\Phi$ in the expanding background. For further simplification, certain steps are proven to be convenient. Considering that the GC is supposed to account for the dark energy $M \sim 10^{-3}$,
and thus we have $\mu \sim H $. This implies that at large distances $\gg \mu^{-1}$ or $H^{-1}$, all terms of \eqref{equation-phi}, except for the last line, are negligible. In the absence of the matter sources,
we therefore find the solutions at fixed large distance $r$ as follows
\begin{equation}\label{phi-large-distance} 
  \Phi \sim c_1(r) e^{- 3 H t} + c_2 (r) e^{ - H t }\,.
\end{equation}
$\Phi$ could be  decomposed into two parts:
\begin{equation}\label{phi-spliting} 
  \Phi = \Phi_\text{N}+\Phi_\text{GC}\, ,
\end{equation}
where $\Phi_\text{GC}$ represents the modification the ghost condensation imparts on the conventional general relativity potential, $\Phi_\text{N}$, which satisfies the Poisson equation
\begin{equation}\label{poisson} 
 \left(\frac{\partial}{a}\right)^2 \Phi_\text{N} = \frac{\rho}{2 M_\text{Pl}^2}\,.
\end{equation}
It should be noted that $\rho$ represents a general matter source. Following this decomposition, the Eq. \eqref{equation-phi} becomes
\begin{align}
& \frac{S_1 S_3}{6 S_2 \mpl^2 \mu^2}  \left(\frac{\partial}{a}\right)^6 \Phi_\text{GC} -\frac{S_1 S_3}{2S_2}\frac{H}{\mpl^2 \mu^2}\left(\frac{\partial}{a}\right)^4 \dot{\Phi}_\text{GC} +
\left[ \frac{2S_1}{3S_2 H^2} + \frac{ S_3 H^2}{6 \mpl^2 \mu^2} \left(\frac{3 S_1}{S_2} - 1\right) -\frac{S_4}{6 \mpl^2}  \right]\left(\frac{\partial}{a}\right)^4 \Phi_\text{GC} \nonumber \\ &
+ \left[\frac{S_3 H^4}{2\mpl^2\mu^2} - \frac{2S_1}{3S_2} \frac{\mu^2}{H^2} - \left( 1 + \frac{S_1}{S_2} \right)  \right] \left(\frac{\partial}{a}\right)^2 \Phi_\text{GC}  +\left[\frac{S_3 H^3}{2\mpl^2\mu^2}
- \frac{1}{3H} \right] \left(\frac{\partial}{a}\right)^2 \dot{\Phi} + 3 H^2 \Phi_\text{GC} +  4 H \dot{\Phi}_\text{GC} + \ddot{\Phi}_\text{GC} \nonumber \\ &  \quad = - \left[\frac{S_3 H^3}{2\mpl^2\mu^2}
- \frac{1}{3H} \right]\left(\frac{\partial}{a}\right)^2 \dot{\Phi}_\text{N} - \left[\frac{S_3 H^4}{2\mpl^2\mu^2} - \frac{2S_1}{3S_2} \frac{\mu^2}{H^2} - \left( 1 + \frac{S_1}{S_2} \right)  \right]
\left(\frac{\partial}{a}\right)^2 \Phi_\text{N} \nonumber \\ & \quad = \left[ \left(\frac{S_3}{2}\frac{H^3}{\mpl^2\mu^2} - \frac{1}{3H} \right)\partial_t  + \left(\frac{S_3}{2} \frac{H^4}{\mpl^2\mu^2} -
\frac{2}{3} \frac{S_1}{S_2} \frac{\mu^2}{H^2} - \left( 1 + \frac{S_1}{S_2} \right)  \right)  \right]  \frac{\rho}{2 M_\text{Pl}^2}.\label{modification-phi} 
\end{align}
A notable issue in the above equation is that the differential equation for $\Phi_\text{GC}$ explicitly depends on the the space and time variation of matter density. We would like to obtain the potential as a function of physical (spherical) coordinate
$r = a(t) x$
\begin{align}  \label{physical-phi} 
 & \partial^6_r \Phi_\text{GC} + 3\Big(\frac{2}{r} - H^2 r\Big) \partial^5_r \Phi_\text{GC} - 3H^2 \partial_r^4 \dot{\Phi}_\text{GC}  - \Big( \frac{S_2 S_4 \mu^2 }{S_1 S_3} - \frac{4 \mu^2 \mpl^2}{S_3 H^2} + \frac{S_2 H^2}{S_1} + 21 H^2 \Big)\partial^4_r \Phi_\text{GC} \nonumber \\ &  - \frac{12 H^2}{r} \partial_r^3 \dot{\Phi}_\text{GC} - \Big(\frac{1}{r}\Big(\frac{4 \mu^2 S_2 S_4}{S_1 S_3} - \frac{16 \mu^2 \mpl^2}{S_3 H^2} + \frac{4 H^2 S_2}{S_1}  + 24 H^2 \Big) + r \Big( \frac{2 S_2 \mu^2 \mpl^2}{S_1 S_3}-\frac{3 S_2 H^4}{S_1}  \Big) \Big) \partial^3_r \Phi_\text{GC} \nonumber \\ & + \Big( \frac{3 S_2 H^3}{S_1} - \frac{2 S_2 \mu^2 \mpl^2 }{S_1 S_3 H} \Big) \partial_r^2 \dot{\Phi}_\text{GC} + \Big( \frac{6 S_2 H^2 \mu^2 \mpl^2}{S_1 S_3} r^2  + \frac{15 S_2 H^4}{S_1} - \frac{24 \mu^2 \mpl^2}{S_3} + \frac{4 \mu^2 \mpl^2}{S_3 H^2} - \frac{14 S_2 \mu^2 \mpl^2}{S_1 S_3} \Big) \partial_r^2 \Phi_\text{GC} \nonumber \\ & + \frac{1}{r} \Big(\frac{6 S_2H^3}{S_1} - \frac{4 S_2 \mu^2 \mpl^2}{S_1 S_3 H} \Big) \partial_r \dot{\Phi}_\text{GC} + \Bigg(\frac{4}{r} \Big( \frac{3 S_2 H^4}{S_1} - \frac{12 \mu^2 \mpl^2}{S_3} + \frac{2 \mu^4 \mpl^2}{S_3 H^2} - \frac{4S_2 \mu^2 \mpl^2}{S_1 S_3} \Big) \nonumber \\ &  + \frac{15 S_2 H^2 \mu^2 \mpl^2}{S_1 S_3} r \Bigg) \partial_r \Phi_\text{GC} + \frac{6 S_2 \mu^2 \mpl^2 }{S_1 S_3} \Phi_\text{GC} \left(3 H^2 \Phi_\text{GC} +  4 H \dot{\Phi}_\text{GC} + \ddot{\Phi}_\text{GC}\right) \nonumber \\ & \quad = \left[\frac{S_2}{S_1 S_3} \Big( 3 S_3 H^3 -\frac{ 2 \mu^2 \mpl^2}{H} \Big)(\partial_t + H r \partial_r)  + 3\frac{S_2}{S_1}\Big( H^4 - 2 S_2 \mu^2 \mpl^2\Big) + \frac{2 \mu^2 \mpl^2}{S_3} \Big(\frac{2 \mu^2}{H^2} - 3 \Big) \right]  \frac{\rho}{2 M_\text{Pl}^2}\,.
\end{align}
We can solve the Eq. \eqref{physical-phi} assuming that $S_i$'s have the same order of magnitude. We arrange the coefficients $S_i$ such that they scale as follows:
\begin{align}\label{Si-redefinition} 
    S_{1} \rightarrow S\,, \ \ S_2 \rightarrow \frac{5}{3}S\,, \ \
    S_3 \rightarrow 4 S\,, \ \ S_4 \rightarrow \frac{11}{3}S\,.
\end{align}

Introducing dimensionless length $X =\mu \ r$ and dimensionless time $T = \Upsilon t$, we have
\begin{align}
& \frac{27}{5 \beta^2} \partial_X^6 \Phi_\text{GC} + \left(\frac{162}{5 X}  - \frac{81}{5 \alpha^2 \beta^2} X\right) \partial_X^5 \Phi_\text{GC} - \frac{81}{\alpha^2 \beta^2} \partial_X^4 \partial_T \Phi_\text{GC} +
\left( \frac{2 \alpha^4}{5}-\frac{612}{5 \alpha^2 \beta ^2}-\frac{33}{4 \beta^2} \right) \partial_X^4 \Phi_\text{GC} \nonumber \\ & -\frac{324}{5 \alpha^2 \beta^3 X}\partial_X^3 \partial_T \Phi_\text{GC} +
\left(\frac{27  X}{\alpha ^4 \beta ^2} -\frac{\alpha ^2  X}{3} +\frac{8 \alpha ^4}{5  X} -\frac{828}{5 \alpha ^2 \beta ^2  X}-\frac{33}{\beta ^2  X} \right) \partial_X^3 \Phi_\text{GC} + \left( \frac{27}{\alpha ^4 \beta ^3}-
\frac{\alpha ^2}{3 \beta }\right) \partial_X^2 \partial_T \Phi_\text{GC} \nonumber \\ & + \left( \frac{135}{\alpha ^4 \beta ^2}+\frac{2 \alpha ^4}{5}-\frac{71 \alpha ^2}{15}+X^2 \right) \partial_X^2 \Phi_\text{GC} +
\left( \frac{54}{\alpha ^4 \beta ^3 X}-\frac{2}{3 \beta  H^3 X} \right) \partial_X \partial_T \Phi_\text{GC} \nonumber \\ & + \left( \frac{108}{\alpha ^4 \beta ^2 X}+\frac{4 \alpha ^4}{5 X}-\frac{112 \alpha ^2}{15 X}+5 X \right)
 \partial_X \Phi_\text{GC} + \frac{1}{\beta^2} \partial_T^2 \Phi_\text{GC} + \frac{4}{\beta} \partial_T \Phi_\text{GC}  + 3 \Phi_\text{GC} \nonumber \\ & \quad = \left(\frac{\alpha^2}{3} - \frac{27}{ \alpha^4 \beta^2}  \right)
 \frac{X \partial_X \rho}{2 \mu^2 M_\text{Pl}^2} + \left(\frac{\alpha^2}{3 \beta} - \frac{27}{ \alpha^4 \beta^3}  \right) \frac{\partial_T \rho}{2 \mu^2 M_\text{Pl}^2} + \left(\frac{17}{5} \alpha^2 - 2 \alpha^4 -
 \frac{27}{\alpha^4 \beta^2}  \right) \frac{\rho}{2 \mu^2 M_\text{Pl}^2}\,,\label{dimensionless-eqPhi1} 
\end{align}
\begin{figure*}[t!]
\centering
\includegraphics[width=1.0\textwidth]{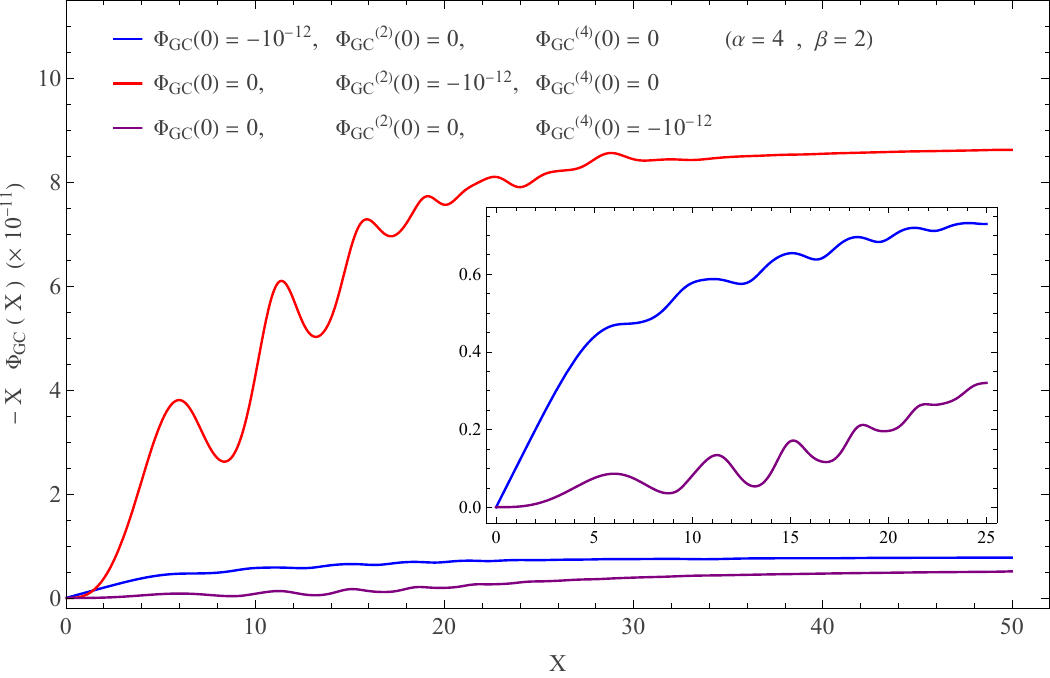}
\caption{For three different boundary conditions, numerical solutions for the equation $\Phi$ are plotted in the form of ($-X \Phi_\text{GC}$) in the absence of a matter source ($\rho = 0$) while $\alpha=4$ and $\beta=2$.}
\label{WithoutMatter}
\end{figure*}
where $\alpha = \mu / H$ and $\beta = H / \Upsilon$. This equation shows that once the ghost with six order dispersion relation is coupled to gravity in the cosmological background, the space and time dependence of the matter source explicitly enters in the EOM for $\Phi_\text{GC}$. In this work we are interested in the modifications that appear at a very late time, i.e. we require the time-independent solution for which $\partial_T = 0$
\begin{align}
& \frac{27}{5 \beta^2} \partial_X^6 \Phi_\text{GC} + \left(\frac{162}{5 X}  - \frac{81}{5 \alpha^2 \beta^2} X\right) \partial_X^5 \Phi_\text{GC} + \left( \frac{2 \alpha^4}{5}-\frac{612}{5 \alpha^2 \beta ^2}-\frac{33}{4 \beta^2} \right) \partial_X^4 \Phi_\text{GC} \nonumber \\ & + \left(\frac{27  X}{\alpha ^4 \beta ^2} -\frac{\alpha ^2  X}{3} +\frac{8 \alpha ^4}{5  X} -\frac{828}{5 \alpha ^2 \beta ^2  X}-\frac{33}{\beta ^2  X} \right) \partial_X^3 \Phi_\text{GC} + \left( \frac{135}{\alpha ^4 \beta ^2}+\frac{2 \alpha ^4}{5}-\frac{71 \alpha ^2}{15}+X^2 \right) \partial_X^2 \Phi_\text{GC} \nonumber \\ & + \left( \frac{108}{\alpha ^4 \beta ^2 X}+\frac{4 \alpha ^4}{5 X}-\frac{112 \alpha ^2}{15 X}+5 X \right) \partial_X \Phi_\text{GC} + 3 \Phi_\text{GC} \nonumber \\ & \quad = \left(\frac{\alpha^2}{3} - \frac{27}{ \alpha^4 \beta^2}  \right) \frac{X \partial_X \rho}{2 \mu^2 M_\text{Pl}^2} + \left(\frac{17}{5} \alpha^2 - 2 \alpha^4 - \frac{27}{\alpha^4 \beta^2}  \right) \frac{\rho}{2 \mu^2 M_\text{Pl}^2}\,.\label{time-independent-eqPhi1}
\end{align}
Here, the elimination of instability is dependent on the sign of the coefficients. Two cases are considered to solve the Eq. \eqref{dimensionless-eqPhi1}
\begin{enumerate}
    \item Without matter source ($\rho = 0$),
    \item  With sources ($\rho \neq 0$).
\end{enumerate}
For the case $\rho = 0$, the equation is numerically solved and three independent solutions are considered. They are shown in Fig. \ref{WithoutMatter}.
Focusing mainly on the overall form of the potential, it can be observed that the Newtonian potential acquires oscillatory behavior which gradually disappears at large scales. At large distances, it approximately approaches  the $R^{-1}$ behavior.
\begin{figure*}[t!]
\centering
\includegraphics[width=1.0\textwidth]{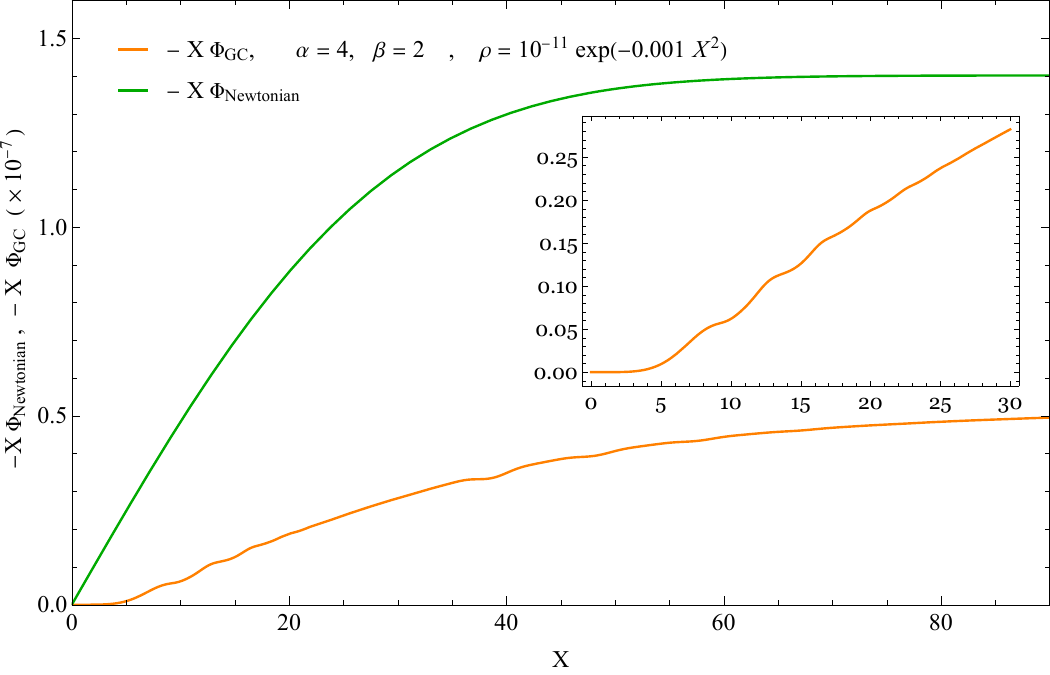}
\caption{The numerical solution with a matter source, $(\rho/\mu^2 = 10^{-11}\exp(-10^{-3} X^2))$, is obtained for the boundary conditions $(\Phi_{\text{GC}}(0) = 0, \ \partial_X^2 \Phi_{\text{GC}}(0) = 0)$, with $\alpha = 4, \beta = 2$.}
\label{Matter}
\end{figure*}
A solution with $\rho = 10^{-11} \mu^2 \exp(- 10^{-3} X^2)$ for $\alpha = 4 $, and $\beta =2$ is depicted in Fig. \ref{Matter} alongside the Newtonian potential. Similar disappearing fluctuations are observed in this case too.

\section{Tensor Perturbation}

Let us focus on the tensor perturbations of the metric
\begin{equation}\label{tensor-metric}  
  g_{ij} = - 2 a^2(t) ( \delta_{i j} + \mathcal{H}_{i j})\, ,
\end{equation}
where $\mathcal{H}_{i j}$ is transverse and traceless,
\begin{equation}\label{tensor-part} 
   \mathcal{H}_{i i} = 0\, , \quad \quad \partial_i \mathcal{H}_{i j} = 0\, .
\end{equation}
The metric \eqref{tensor-metric} allows us to study tensor perturbations. The expansion of the Einstein-Hilbert action gives
\begin{equation}\label{tensor-EH-action} 
  A_\text{EH} = \frac{\mpl^2}{8} \int \mathrm{d}t \ \mathrm{d}^3 x \ a^3 \left[\left(\dot{\mathcal{H}}_{i j}\right)^2 - a^{-2} \left(\partial_k \mathcal{H}_{i j} \right)^2 \right]\, .
\end{equation}
Operators proportional to $\sigma_2$, $\sigma_3$ and $\sigma_4$ yield no contribution to the second order expansion of the action in terms of $\mathcal{H}_{i j}$. On the other hand, $\gamma^{ij} \nabla_i K_{ml} \nabla_j K^{ml}$
produces the following action at second order
\begin{equation}\label{sigma1-action} 
  A_{\sigma_1} = -\frac{\sigma_1}{2} \int \mathrm{d}t \ \mathrm{d}^3 x \ a^3 \left[ a^{-2} \left(\partial_k \dot{\mathcal{H}}_{i j} \right)^2 \right]\, .
\end{equation}
This means that the speed of tensor perturbations gets a momentum-dependent modification.
Using the conformal time, $d\eta=\frac{dt}{a}$ and $u_k$, the following modified EOM for tensor perturbation is obtained
\begin{equation}\label{eom-tensor} 
  u_k(\eta) \left(\frac{k^2 \mpl^2}{\mpl^2 - 4 \sigma_1 k^2 }-\frac{a''(\eta)}{a(\eta)}\right)+u_k''(\eta) =0\,,
\end{equation}
where $u = a(\eta) \mathcal{H}_k $ and $\mathcal{H}_k$ stands for one of the two Fourier components of the tensor perturbation mode function. This means that the speed of propagation of GWs is given by
\begin{equation}\label{cT} 
c_T^2 = \left(1 - 4\sigma_1 \frac{k^2}{\mpl^2}\right)^{-1}\, .
\end{equation}
We notice that the speed of gravitational waves is momentum dependent. Also $\sigma_1\leq0$ in order for the gravitational waves to be subluminal. For each $k$, deviation from the speed of light becomes substantial if $|\sigma_1|\sim \frac{\mpl^2}{k^2}$. On the other hand, it is expected that $H\lesssim k \lesssim M$, where the latter inequality comes from the viability of the effective field theory. For $k\sim M\sim 10^{-3}$ eV, this means that $\sigma_1$ has to be $\lesssim 10^{60}$ to see substantial deviation from the luminal propagation for the GWs. For $k\sim H \sim 10^{-33}$ eV, this changes to $\sigma_1\lesssim 10^{120}$. Otherwise the deviations of the speed of GWs from the light speed is quite small and negligible. From the LIGO-Virgo observations of the binary neutron star merger GW170817, $-3\times 10^{-15}<c_T-1<7\times 10^{-16}$ in the frequency range $10~ {\rm Hz}\lesssim f \lesssim 10~ {\rm kHz}$ \cite{LIGOScientific:2017zic,LIGOScientific:2017vwq,LIGOScientific:2017ync}, one obtains that $|\sigma_1|\lesssim 10^{69}$ for $f\sim 10~{\rm Hz}$ and $|\sigma_1|\lesssim 10^{63}$ for $f\sim 10~{\rm kHz}$.

\section{Conclusion and Directions for Future Research}

Following our observation in the previous work \cite{Ashoorioon:2023zju} that GC with a sextic dispersion relation can serve as a viable effective field theory without strong coupling in the IR, the role of a GC dark energy with a sixth-order dispersion relation in de Sitter space is explored. Coupling such a ghost condensate model with gravity reveals important modifications of dispersion relation and the propagation speed for the dark energy perturbations. The dispersion relation in addition to the sixth order dispersion relation, obtains both quartic and quadratic terms. As for the gravitational potential, our work demonstrates that contrary to the GC with quartic dispersion relation, its EOM depends explicitly on the space and time variations of the matter source. For modification that appear at very late time, regardless of presence or absence of matter sources, the gravitational potential obtains oscillatory modulations at the scales $\sim \frac{M^2}{\mpl}$ in the timescales of order $\frac{{\mpl}^3}{M^4}$, before it approaches the $R^{-1}$ behavior at large distances.
 In addition, we found the speed of GWs to be momentum dependent where the deviations from the speed of light becomes important at momenta comparable with $\frac{\mpl}{\sqrt{|\sigma_1|}}$, where $\sigma_1$ is the coefficient of $\gamma^{ij} \nabla_i K_{lr} \nabla_j K^{lr}$ operator in the unitary gauge action.

 In the course of computing the equation governing the gravitational potential, we noticed that even in the absence of anisotropic stress tensor, there is a possibility for $\Phi\neq\Psi$. It would be interesting to investigate the observational consequences of this possibility. Also in this work, we computed the late time solutions for $\Phi$ for which $\partial_T=0$. It would be interesting to investigate the effect of time-dependent matter sources in the cosmological setup and find out the consequences it has for the structure formation.

\section*{Acknowledgements}

We are thankful to Roberto Casadio and Paolo Creminelli for helpful discussion. A.A. is thankful to ICTP, where part of this project was accomplished. This project is supported by INSF grant 4031449.
This project has also received funding from the European Union's Horizon Europe research and innovation programme under the Marie Sk\l odowska-Curie Staff Exchange grant agreement No 101086085-ASYMMETRY. 

\bibliographystyle{elsarticle-num}
\bibliography{ref.bib}

\end{document}